%% file: ConceptualModelForHighlights_art.tex
\documentclass{article}

\usepackage[utf8]{inputenc}
\usepackage[english]{babel}
\usepackage{xcolor,colortbl}
\usepackage{url}
\usepackage{graphicx}
\usepackage[figuresleft]{rotating}

\usepackage{typed-checklist} 

\DeclareGraphicsExtensions{.png,.jpg,.pdf}

\newcommand{\silence}[1]{}

\definecolor{orange}{rgb}{0.93, 0.57, 0.13}

\newif\ifDRAFT
\DRAFTfalse       

\ifDRAFT
\newcommand{\isDraft}[1]{#1}
\else
\newcommand{\isDraft}[1]{}
\fi

\ifDRAFT
\def\inlineRem#1{{\noindent\bf\fcolorbox{red}{white}{\color{blue}{$\|$~{#1}~$\|$}}}}
\else
\newcommand{\inlineRem}[1]{}
\fi

\ifDRAFT
\def\rem#1{\colorbox{yellow}{\scriptsize\color{brown} @@} \marginpar{\scriptsize\color{brown} #1}}
\else
\newcommand{\rem}[1]{}
\fi

\ifDRAFT
\def\sideNote#1{\marginpar{\indent \color{gray}{\footnotesize{\emph{#1}}}}}
\else
\newcommand{\sideNote}[1]{}
\fi

\newcommand{\forcepagebreak}{\clearpage \newpage}

\def\archProp{archetype property}
\def\archProps{archetype properties}

\title{A Conceptual Model for Data Storytelling Highlights in Business Intelligence Environments}

\author{
    Panos Vassiliadis  \\
	University of Ioannina, Ioannina 45110, Greece \\
	\textsf{pvassil@cs.uoi.gr}, orcid: \url{0000-0003-0085-6776}
    \and
    Patrick Marcel , Faten El Outa, Veronika Peralta\\
	University of Tours, Blois, France \\
    \textsf{firstname.lastname@univ-tours.fr}\\
    orcid: \url{0000-0003-3171-1174}, \url{0000-0002-8153-4252}, \url{0000-0002-9236-9088}
    \and
    Dimos Gkitsakis \\
	University of Ioannina, Ioannina 45110, Greece \\
	\textsf{dgkits@cs.uoi.gr}, orcid: \url{0009-0006-9559-819X]}
}

\begin{document}
	\maketitle

\begin{abstract}
	We introduce a conceptual model for highlights to support data analysis and storytelling in the domain of Business Intelligence, via the automated extraction, representation, and exploitation of highlights revealing key facts that are hidden in the data with which a data analyst works. The model builds on the concepts of Holistic and Elementary Highlights, along with their context, constituents and interrelationships, whose synergy can identify internal properties, patterns and key facts in a dataset being analyzed.
\end{abstract}

\input{body}

\newcommand{\etalchar}[1]{$^{#1}$}

\end{document}

%% file: body.tex
\section{Introduction}
Data narration is an activity of presenting an audience with findings that are derived from underlying data via a process of data extraction, analysis, verification, and multi-modal presentation that comes via a combination of audiovisual means (e.g., text, charts, voice, imagery,  video to name the most prominent ones). Journalists and data analysts, in business and government, work with data to construct data narratives. The main process relies on working with data that answer analytical questions, and the findings are then presented in a rich, multi-modal way to the audience. 

The related literature and practice of data narration \cite{DBLP:journals/tvcg/SegelH10,DBLP:journals/dagstuhl-reports/CarpendaleDRH16,DBLP:conf/er/OutaFMPV20,DBLP:conf/adbis/OutaMPSCV22} demonstrate an interesting gap: whereas in practice the data journalists pay significant efforts at the data analysis side, the literature is mostly emphasizing the presentation part. We argue that the key concept to bridge the gap is to provide a \textit{structured} answer to analytical questions, in a way that this structure is later exploited to semi-automatically construct the presentation part. Already, several efforts towards automating the results of analytical questions have been made \cite{DBLP:journals/is/GkesoulisVM15,DBLP:conf/sigmod/DingHXZZ19,DBLP:journals/tvcg/WangSZCXMZ20,DBLP:conf/er/OutaFMPV20,DBLP:conf/sigmod/00040HZ21,DBLP:journals/tvcg/ShiXSSC21}: the typical terminology for these automatically extracted answers are \textit{insights}, or, \textit{messages}. An insight/message concerns the parts of the data space that demonstrate properties worth reporting to the audience. When, for example, a time series demonstrates a trend, or a histogram follows a certain distribution, then the data demonstrate a phenomenon that may, or may not, be worth reporting. When certain phenomena in the data are of such importance that they deserve to be reported to the audience, then, we have the emergence of \textit{highlights}, i.e., the structured annotation of the parts of the data that make the existence of an interesting phenomenon true with the necessary information that explains why these data are of importance. The key contribution of this paper is the introduction of a conceptual model for highlights, in the context of business intelligence. But first, we will illustrate the case with an example. 
\sideNote{Highlight covers two of the main concepts of data narrative, which are "finding" and "message." Specifically, it covers "finding," which involves identifying significant discoveries that emerge from the analysis of collected data and represent new insights, patterns, or relationships. It also covers "message," which involves synthesizing and organizing related findings into a clear and concise information that conveys the most important and relevant insights to the intended audience. However, the key distinction of highlight is that it provides more explicit explanation as to why the highlighted data are of importance. This process involves annotating specific parts of the data and providing the necessary context and information to demonstrate why they are relevant to the analysis and the overall narrative.}


\textbf{Working Example}. 
Assume we have a database with product sales. The fundamental source of information is a fact table \textit{SalesFact(\underline{ProductId}, \underline{TimeId}, \underline{CityId}, \underline{PromotionId}, \underline{CustomerId}, Sales, Costs, UnitSales, AvgSalesPerUnit, Profit)}, with all the monetary measures expressed in thousands  of Euros. Around the aforementioned fact table, we also have several lookup dimension tables, namely \textit{Products, Time, Cities, Promotions, Customers}, joinable to the fact table via the respective $Id$ attributes (i.e., we have a PK-FK relationship between fact and dimension tables for each $Id$ attribute). This is as typical a Business Intelligence setup as it gets. \rem{univ. relation, tods 84 maier ulmann}

Assume a query that selects the total sum of sales of the product \textit{Wine} for the 2nd quarter of 2023 in Greece, grouped by month and city. Observe the following \textit{highlights} of the result set:
\begin{itemize}
	\item The city \textit{Athens} dominates all other cities: for every month, the sales of Athens are higher than the sales of every other city.
	\item The month \textit{May 2023} dominates all other months: for every city, the sales of May are higher than the sales of other months.
	\item The city of \textit{Athens} is a mega-contributor to the total sales: the sales of \textit{Athens} are 75\% of the total sales.
	\item If one observes the time-series of the marginal sales per month, there is no trend or seasonality; however there is a unimodality in the time-series: sales rise, reach a peak (in \textit{May}), and then drop.
\end{itemize}

\begin{table}[b]
	\centering
	\begin{tabular}{r|rrrr|r}
		& \textit{Athens} & \textit{Rhodes} & \textit{Chania} & \textit{Thera} &               \\
		\hline
		\textit{April 2023} & 500             & 50              & 85              & 80             & \textit{715}  \\
		\textit{May 2023}   & 1000            & 70              & 90              & 120            & \textit{1280} \\
		\textit{June 2023}  & 600             & 65              & 70              & 70             & \textit{805}  \\
		\hline 
		& \textit{2100}   & \textit{185}    & \textit{245}    & \textit{270}   & \textbf{\textit{2800}}
	\end{tabular}%
	
	\caption{Reference Example}
	\label{tab:refEx}
\end{table}

How do we report the results to the user? With the current tools of the trade, the only thing to be reported is the result of the query, as returned by the DBMS, possibly accompanied by the marginal sums, which can easily be computed client-side from a dedicated application. What we want is a system that derives a \textit{data narration} that includes (a) the data per se (as E.Tufte has famously  emphasized ``always show the data"), (b) appropriate visualizations that are automatically produced, and, (c) text narrating the essential elements of the data. 

The following textual summary reports all the discovered highlights, and groups them around the main \textit{characters} of Athens and May: 
\begin{quote}
	\textit{In terms of geography, Athens dominates all other cities, in every month. In fact, Athens is a mega-contributor to total sales, by contributing 75\% of all sales. In terms of time, the progression in time shows a peak in May; in fact, May dominates all other months in terms of total sales. No trend or seasonality were detected.}
\end{quote}

For the interest of space, we have omitted the visualizations here and focused only on the textual narration. However, the same idea applies to visualizations too: dominators are depicted via horizontal bar charts; time series via line charts; important characters or values appropriately highlighted with color.

\textit{The main contribution of this paper lies in the provision of a comprehensive and precise conceptual model for highlights, along with their constituents and their interrelationships, in the domain of Business Intelligence, with the goal to help system builders implement tools and algorithms that facilitate the automated extraction, representation, and exploitation of highlights in data, for data analysis and storytelling purposes}. 

Specifically, in Section \ref{sec:RW}, we survey related work. In Section \ref{sec:model}, we introduce the model for highlights. We separate the model in (a) a model for back-end data, practically introducing the main concepts of \textit{Character} and \textit{Measure Value}, and (b) front-end highlights, discriminating highlights in \textit{Holistic Highlights}, which are properties of the entire dataset being examined, and \textit{Elementary Highlights} which concern individual \textit{Characters}, or sets of them, that play a crucial role to \textit{Holistic Highlights}. In Section \ref{sec:discuss}, we discuss the practical usage of the proposed model. We conclude our discussion in Section \ref{sec:fw} with a summary and issues for future work.

\section{Related Work}\label{sec:RW}
In this Section, we review the related work around the concept of highlights, as well as related areas.

\textbf{Data exploration and OLAP.} 
Exploratory Data Analysis (EDA), the notoriously tedious task of interactively analyzing datasets to gain insights, has attracted a lot of attention lately \cite{DBLP:conf/sigmod/IdreosPC15}.
Supporting this task can be done, e.g., by generating EDA notebooks using deep learning \cite{DBLP:conf/sigmod/ElMS20} which supposes having access to lots of former analyses, or by pre-analyzing datasets for computing so-called insights \cite{DBLP:conf/sigmod/TangHYDZ17}.
EDA is similar to \emph{Discovery-Driven Exploration} (DDE) of data cubes \cite{DBLP:conf/edbt/SarawagiAM98}, 
essentially motivated by explaining unexpected data in the result of a cube query. Unexpectedness was characterized in terms of deviation from the uniform distribution  \cite{DBLP:conf/vldb/Sarawagi00}, or notable discrepancies in the data to be explained by generalization (rolling-up) \cite{DBLP:conf/vldb/SatheS01}, or by detailing (drilling-down) \cite{DBLP:conf/vldb/Sarawagi99}.
Gkesoulis et al. \cite{DBLP:journals/is/GkesoulisVM15} demonstrated how to enrich query answering with a short data movie that gives insights to the original results of an OLAP query. 

\textbf{Defining highlights.}
There is no clear consensus on the terminology used for the case of highlights. The term insights is well adopted in the data management and data visualization communities \cite{DBLP:conf/sigmod/DingHXZZ19,DBLP:conf/sigmod/00040HZ21}, and other terms are also used, like discoveries \cite{DBLP:conf/edbt/SarawagiAM98} or data facts \cite{DBLP:journals/tvcg/WangSZCXMZ20}. \textit{We believe that a clarification of the related concepts is one of the contributions of this paper}.
Various definitions of highlights are independently proposed in the literature.
In Datashot \cite{DBLP:journals/tvcg/WangSZCXMZ20}, a so-called data fact is a tuple composed of 
(i) type: type of information described (e.g., outlier, difference, extreme),
(ii) parameter: characteristic of the type (e.g., min or max for extreme),
(iii) measure: the dependent variable for the fact (e.g., sales),
(iv) subject: a set of filters of the form \textit{dimension} = \textit{value} over a dataset, and, 
(v) score: the importance of the fact for a user.
This definition is refined in \cite{DBLP:journals/tvcg/ShiXSSC21}, where
the subject corresponds to a subspace (filters) with a breakdown
(categorical attribute used to divide the subspace into groups)
and a focus (part of the subspace deserving attention).

In \cite{DBLP:conf/sigmod/DingHXZZ19,DBLP:conf/sigmod/00040HZ21}, the authors define
insights and what they call meta-insight. A data-scope is a triple composed of a subspace, a measure and a grouping dimension. Data patterns types correspond to data fact types (outlier, extreme, etc.). A data pattern (insight) is a data-scope with a type and a parameter (type dependent characteristic of the data). Meta-insights are sets of commonness (homogeneous insights in the sense of some similarity function) and exceptions (to the insights).

Importantly, all these definitions consider a multidimensional dataset.

\textbf{Data narration.}
Chen et al. \cite{DBLP:journals/tvcg/ChenLAAWNT20} emphasized that in order to create a compelling data story, it is crucial to convey well-organized and assembled information pieces rather than solely focusing on the findings obtained from the analysis.
Outa et al. \cite{DBLP:conf/er/OutaFMPV20,DBLP:conf/adbis/OutaMPSCV22} proposed a conceptual model for data narratives by providing a principled definition of the key concepts of the domain, along with their relationships. Among these concepts, they introduced the concept of ``Message" (a partial answer to an analytical question), which is reliant on another concept called ``Finding" (a significant discovery that emerges from the analysis of collected data), via a N:M relationship.

\textbf{Interestingness of highlights.}
Characterizing meaningful highlights in data has attracted a lot of attention, from the seminal work on discovery driven exploration \cite{DBLP:conf/edbt/SarawagiAM98,DBLP:conf/vldb/Sarawagi00} and on knowledge discovery in databases \cite{DBLP:conf/vldb/AgrawalS94}.
Often, this characterization takes the form of \textit{interestingness} scores for retrieved data \cite{DBLP:conf/adbis/MarcelPV19} or \textit{patterns} \cite{DBLP:journals/csur/GengH06,DBLP:conf/ida/Bie13}.
Scoring a finding is used to express its importance, or interestingness, for the user. As explained in \cite{DBLP:conf/adbis/MarcelPV19,DBLP:conf/sigmod/MiloS20},  interestingness is manifold: scores can be computed for different dimensions of interestingness. In the taxonomy proposed in \cite{marcel2023data}, following \cite{DBLP:conf/sigmod/PatilAS22},  interestingness of findings can be characterized with human, system or data metrics.
Chen et al. \cite{DBLP:journals/tvcg/ChenLAAWNT20} stress the importance of the relationships between findings for the selection of the more important ones, while other authors exploit their \textit{significance}, defined in terms of statistical tests \cite{DBLP:journals/tvcg/GuoGZL16,DBLP:conf/sigmod/TangHYDZ17,DBLP:conf/chi/ZgraggenZZK18,DBLP:journals/tkde/JoglekarGP19,DBLP:journals/is/GiuzioMQRST19,DBLP:journals/tvcg/WangSZCXMZ20}, Shapley values \cite{DBLP:journals/pvldb/DeutchGMS20}, or information theory \cite{DBLP:journals/tvcg/ShiXSSC21}.

\section{A model for highlights}\label{sec:model}
In this Section, we present the model for highlights, along with its supportive concepts.

\subsection{A model for highlights}
Given a dataset, which can possibly be the result of a query, highlights are important properties of the entire dataset, or individual facts and characters that are produced via the analysis of the dataset. 
\isDraft{We can discriminate highlights in two kinds:
	\begin{itemize}
		\item \textit{Manually constructed highlights}, that the storyteller puts down as notes or observations for subsequent exploitation.
		\item \textit{Automatically extracted highlights}, to which we will also simply refer as \textit{highlights} (as they are the essence of our discourse), that a data storytelling system can \textit{algorithmically} extract from the underlying data, in order to support the data exploration and storytelling process.
	\end{itemize}
	
	\subsubsection{Phenomena and Highlights}
	\paragraph{What is to be reported?} The previous discussion begs the question on defining what is (a) useful for reporting purposes, and, at the same time, (b) algorithmically feasible, in order to be classified as a highlight. We will first introduce the process of highlight extraction, and then move on to provide the necessary definitions. 
	
} 

\begin{figure}[bt]
	\includegraphics[width=\columnwidth]{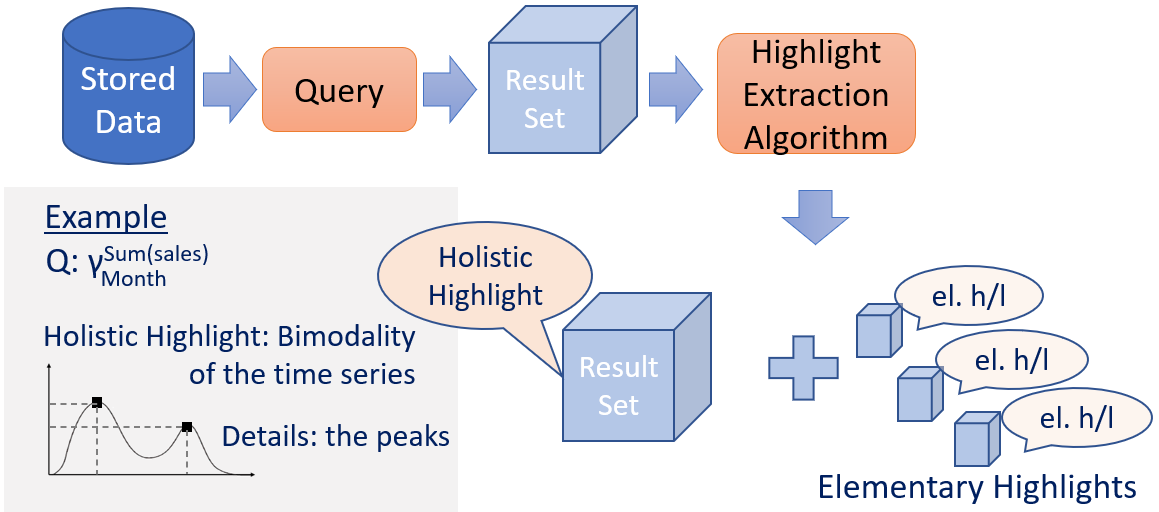}
	\caption{The process of highlight extraction} \label{fig:procHL}
\end{figure}

The process of automated highlight extraction (Figure \ref{fig:procHL}) practically applies a set of Highlight Extraction Algorithms over a dataset. These algorithms operate like pattern-matching testers: they try to see whether the data abide by a certain pattern or not.  To give a couple of concrete examples, possible questions that such  algorithms might ask are: (a) is there a bimodality in a time-series produced as a query result, and if yes, which are the peaks?, (b) if we breakdown the total sum of sales by product type, is there a ``mega-contributor" product type (and if yes, who is it), with more than 40\% of total sales?, (c) assuming we group by total sales per month and product is there any month that systematically outperforms all other months for all products, and if yes, who is it? 
\sideNote{Is every phenomenon applied to the data, or is there a process for selecting which phenomenons to use? In cases where there are multiple pattern matches, how is the appropriate one chosen? 
	Questions are not detailed. How the questions are defined? Does it rely on the features of the data?}

\subsubsection{Supporting concepts} 
We assume a finite set of \archProps\ (like, e.g., a trend in a timeseries, the correlation of two Measure Types, the existence of mega-contributors, etc.) with well known semantics. 

\textbf{Definition}. An \textbf{\archProp}\ is a specific pattern of internal relationships between the contents of a dataset. 

For a dataset to support an \archProp\, we require that when a verifying algorithm is applied to it, it can produce a meaningful resulting \textit{Model} that tells us whether the \archProp\ exists or not. This result, can, for example, involve the existence of one or two peaks in a time series for uni/bi-modality, the percentage of a Character's contribution to a total sum for mega-contributors, the p-value for a distribution test, the MSE for a regression formula, the correlation score for a correlation test, etc. Moreover, in some cases, like for instance in the case of modality peaks, top-k values, or peer-domination, it is possible that there exist specific \textit{Characters}, \textit{Character} combinations and \textit{Facts} that play specific roles to facilitate the verification of the \archProp. These are the details of the existence of the \archProp\, without which the reporting of the \archProp\ is useless (you need to tell the analyst at which timepoint the time series reaches a peak, for example). In other cases, e.g., in the cases of a distribution check, or the correlation of two Measure Types, these details are not present.

\begin{figure*}[tb]
	\centering
        \includegraphics[height=0.53\textheight]{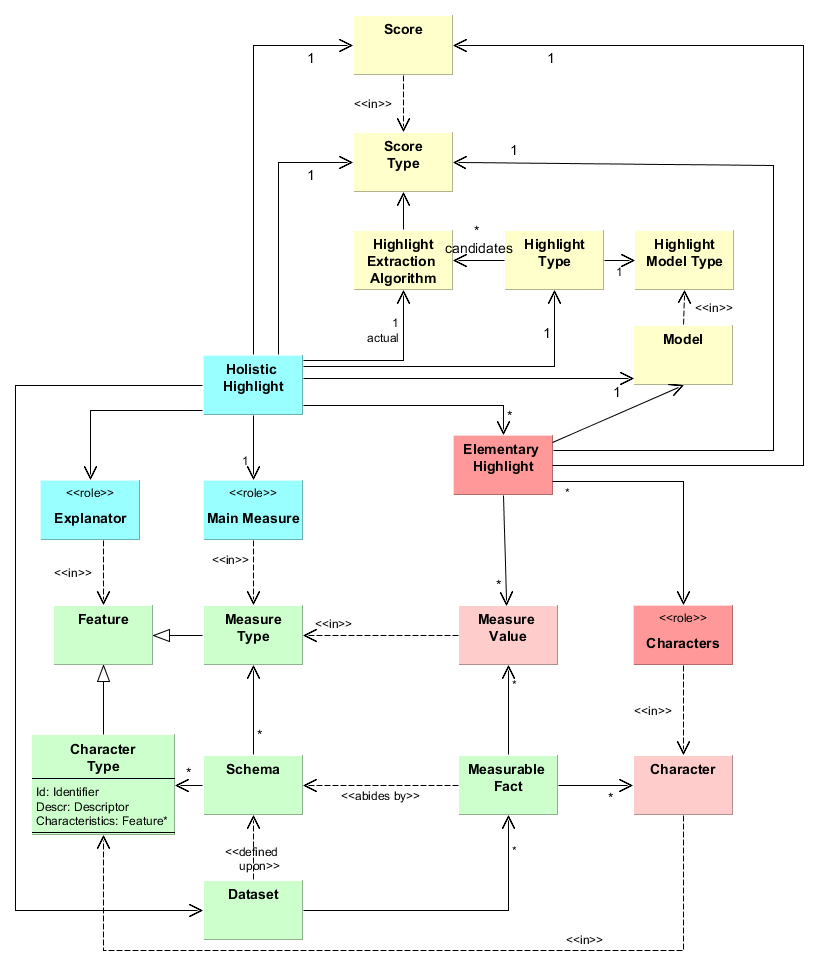}

	\caption{The metamodel for highlights} \label{fig:hlModel}
\end{figure*}

Data are recorded as \textit{facts}, abiding by the structure induced by a \textit{Schema} (aka \textit{Fact Type}). A fact is an  observation for the domain being modeled. Schemata define a template internal structure for the facts. \textit{Features} are single-valued attributes with the typical data types used  in databases and programming languages, which, for practical purposes, we coarsely classify in four subclasses: \textit{Identifiers} (used to uniquely identify composite structures), text \textit{Descriptors}, as well as \textit{Numeric} and \textit{DateTime} \textit{Features}. Each feature has a domain of values. A \textit{Dataset} is a set of facts under a \textit{Schemata}. The Cartesian Product of the domains of the Features of the Schema produce the domain for the facts. 


A \textit{Measure Type} is a named domain of quantitative measurements, typically of numeric form (Boolean Measure Types can are also be allowed). For example, a Measure Type like $TotalSales$ can characterize the amount of money made by the sales of products in a store at a given date. \textit{Measure Types} come with \textit{Units}, like kilometers, miles, Celsius degrees -- or just the plain quantity of an item set. A \textit{Measure Value} is an instance of a Measure Type. Whenever a query applies an aggregation function over the underlying data, the resulting schema includes an \textit{Aggregate Measure Type}, produced via an \textit{Aggregate Function}. 
Whenever Measure Types are combined via some algebraic formula (e.g., $Profit$ = $Earnings$ - $Cost$), then the result is a \textit{Derived Measure Type}.

\textit{Characters} are the main entities of the domain being modeled (which in the OLAP terminology would be named as \textit{Dimension Members}). Examples include specific cities (e.g., \textit{Athens}), dates (e.g., \textit{May 2023}), products, etc. A \textit{Character Type} is a structured named domain of \textit{Characters}.  A \textit{Character Type} comes with the following three kinds of characteristics: (a) an $Id$ that uniquely identifies each Character, (b) a $Description$ that provides a textual, human-relatable description of the Character, and (c) a set of $Characteristic~Properties$, that each Character Type carries with it. For example, for the Character Type $City$, one can envision the schema $City(Id, Description, Area, Population)$. A possible character would be \textit{AthensMetropolitan} \textlangle \textit{101, ``Athens Metropolitan Area", 2928} $Km^2$, \textit{3.7M} \textrangle. \rem{PM: Can we have Dimensions and Levels as characters themselves?}

\subsubsection{Holistic highlights}

Holistic highlights characterize properties of an entire dataset (typically, but not necessarily, the result set of a query). The emphasis is not on specific data points of the result set, but rather on the dataset as a whole. Sometimes, the \archProp\ that the highlight captures also distinguishes individual facts, as details of the holistic highlight. However, the characterization refers to the entire result set. Let us begin with a couple of examples to illustrate the idea, and then move on to present the specificities of holistic highlights.
\begin{enumerate}
	\item \textsf{Distribution.} The \textsf{distribution of the values} of a certain Measure Type, say $M$, follows the \textsf{Normal distribution} model. The test was performed by a Shapiro-Wilk test and the p-value is $10^{-4}$. 
	\item \textsf{Correlation.} The \textsf{correlation} of a  Measure Type, say $M$, with another Measure Type $M'$, is characterized as \textsf{significant}. The test was performed via a Kendall test and the t-value is $0.83$.
\end{enumerate}

\textbf{Definition}. A \textbf{Holistic Highlight} is a significant, structured testimony for the existence of an \archProp\ over a specific dataset that is automatically tested via a dedicated algorithm and characterized accordingly. Structurally, a holistic highlight is characterized by the following properties:

	The \textit{Highlight Type} defines the family to which the highlight belongs (e.g., Correlation, Trend, Seasonality, Unimodality, Bimodality, etc.)
	
	The \textit{Highlight Model Type} defines the domain of results that the test for an \archProp\  can produce. For example, when examining the distribution of the values of a Measure Type, the model can be any of \textit{Normal}, \textit{Powerlaw}, \textit{Uniform}, etc. When assessing the correlation of two Measure Types the Model can be any of \textit{Positively Significant}, \textit{Insignificant}, \textit{Moderately Negatively Significant}, etc.
	
	The \textit{Model} of the highlight defines the state of the data with respect to the Highlight Type under investigation. Practically, this is achieved via the specific value from the domain of the Model Type that pertains to the result set being characterized.
	
	The test of the \archProp\ is performed via one of the candidate algorithms that exist for the same Highlight Type. For example, Correlation has several candidate algorithms: Pearson, Spearman, Kendall, etc. However, every highlight is produced via the execution of a specific algorithm out of the set of candidate ones, which we refer to as \textit{Actual Algorithm}. The actual algorithm uniquely determines the Model Type of the highlight. 
	
	Moreover, the strength of the \archProp\ is characterized by a \textit{Score Type} and an actual \textit{Score}. The \textit{Score Type} is the domain for the actual score -- e.g., a Score Type \textit{p-value} can have an actual \textit{Score} value of 0.3. \rem{PM: is this peculiarity? PV: This is more technical, finding-side stuff: e.g., rank in topK, pct of dominated peers, etc.}
	
	The highlight pertains to a \textit{Measure Type} of the result set that is tested for the existence of an \archProp. For example, the existence of a Bimodality is tested over a certain Measure Type, say $SumSales$.
	
	The  Measure Type alone is not sufficient to give all the information of a highlight. For example, unimodality of a measure has to be produced with respect to a certain time attribute (in a typical BI dataset, a basic cube can have several such attributes, $OrderDate$, $DispatchDate$, $ArrivalDate$ -- therefore the testing of a modality needs to be done with respect to one of them). Thus, a set of \textit{Features} as \textit{Supportive Explanators} is used to accompany the main measure of the highlight in order to fully specify the existence of an \archProp.
	
	Last but not least, a holistic highlight can include a set of details in the form of Elementary Highlights (to be discussed in the sequel).

To be able to discriminate which Character plays which role in a Highlight, we formally introduce Roles in the model. A \textit{Role} annotates a class with specific attributes, specifically: (a) a \textit{name}, (b) an accompanying \textit{textual description}. The \textit{Measure Type} and the \textit{Supportive Explanators} are incarnations of roles. Their participation in the highlight is annotated by textual information that describes their role and the respective Feature that incarnates the role.

A possible textual description of a Holistic Highlight is as follows:

\textit{	The $<HighlightType>$ for $<MainMeasure>$, tested via $<Algorithm>$, $\{SupportiveRoleText; SupportiveRole\}\star$, fits under the $<Model>$ with $<ScoreType>$ and value $<ScoreValue>$. } 

\subsubsection{Elementary Highlights}

The \textit{Elementary Highlights} are specific facts that contribute with specific roles in a highlight. The fundamental difference from \textit{Holistic Highlights} is that instead of a \textit{Measure} (which is a Feature), the protagonist of the highlight is a \textit{Character} who demonstrates a behavior measured by a \textit{Measure Value}. Two illustrative examples:
\begin{itemize}
	\item \textsf{Top-k} A fact can be in the \textsf{top-k} facts of a query result with respect to a certain \textit{Measure Type}, having a \textit{Measure Value} $v$. Its \textsf{rank} denotes how high in the list of top-k facts, this particular fact is.
	\item \textsf{Unimodality peak} If the dataset presents a unimodality pattern of certain measure over a certain \textit{Character Type}, this means that there is a certain fact describing the \textsf{peak of the distribution of values}. The \textsf{Character acting as the ``x-axis"} \textsf{coordinate of this particular fact} is an Elementary Highlight for the dataset.\\
\end{itemize}

\forcepagebreak  

\begin{sidewaysfigure*}[!ht]
	\includegraphics[width=0.9\columnwidth]{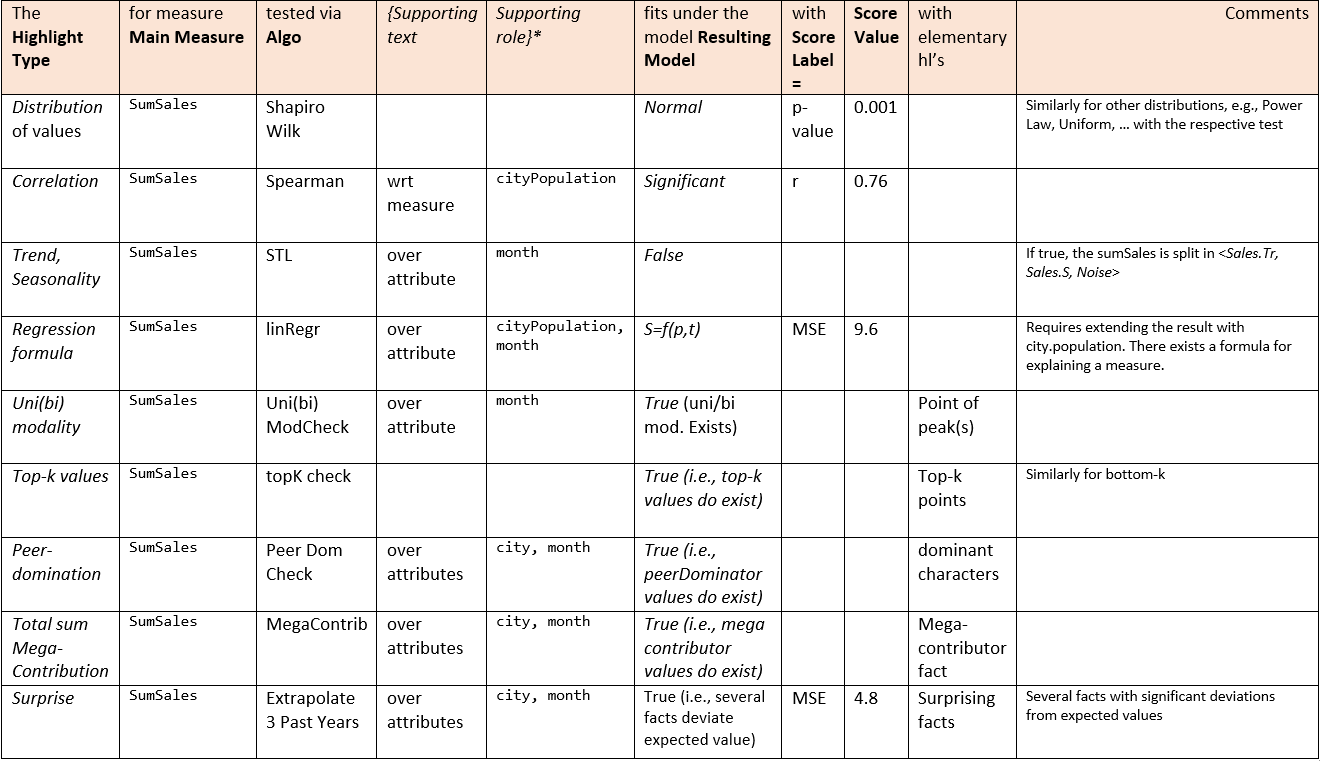}
	\caption{Examples of holistic highlights} \label{fig:exTable01}
\end{sidewaysfigure*}
\forcepagebreak
\begin{sidewaysfigure*}[!ht]
	\includegraphics[width=0.9\columnwidth]{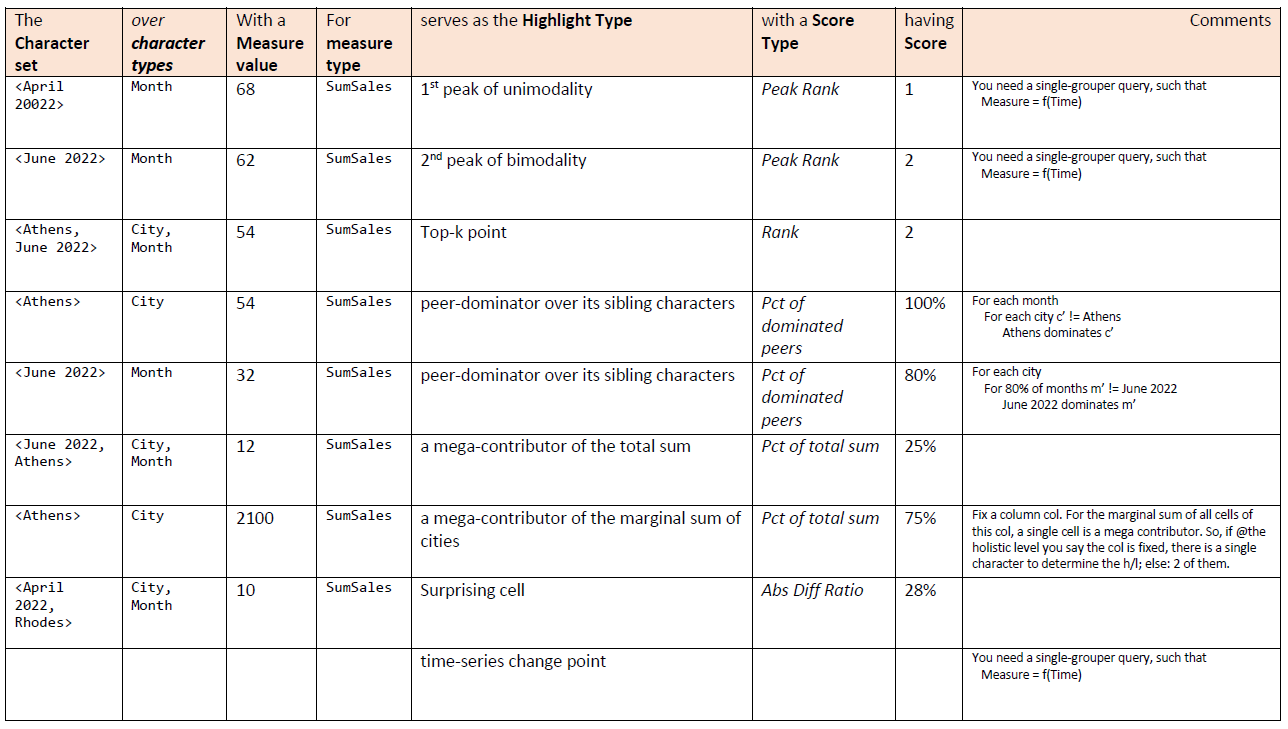}
	\caption{Examples of elementary highlights} \label{fig:exTable02}
\end{sidewaysfigure*}
\forcepagebreak


\textbf{Definition}. An \textbf{Elementary Highlight} is a fact produced by the combination of specific Characters and Measure Values that play an important role in a holistic highlight. Structurally, an Elementary Highlight is characterized by the following properties:

The \textit{Highlight Type} defines the family to which the highlight belongs (e.g., a peak in a unimodality distribution, a part of the top-k facts set for a measure, a mega-contributor fact over its peer facts for the breakdown of an aggregate value, etc.)

The \textit{Character Set} of an Elementary Highlight is a set of identifying  \textit{Highlight Characters}. A \textit{Highlight Character} is a Role, carrying a distinctive name and text that characterize its functionality with respect to the highlight, as well as a specific \textit{Character} that pertains to the highlight. For example, if a query asks for sum of sales per month and city and $<Athens, April~2022>$ are the coordinates of the top-1 fact, then, both \textit{Athens} and \textit{April 2022} participate in the highlight with their own Highlight Character, and, it is their combination that uniquely identifies the elementary highlight. The \textit{Highlight Character Type} defines the type of the each of the \textit{Characters} participating in the \textit{Character-Set} of the highlight.

The \textit{Measure Value} gives the specifics of the Elementary Highlight for a \textit{Measure Type}.
In the aforementioned example, the specific \textit{Measure Value} for the \textit{sum of sales} in \textit{Athens} in \textit{April 2022} gives this information. 
	
As with \textit{Holistic Highlights}, \textit{Elementary Highlights} are also related to the execution of the actual algorithm that produced them via the respective \textit{Model} for which they perform an illuminating role.

Similarly, each Elementary Highlight has a \textit{Score Type} and an actual \textit{Score}, to characterize the strength of the highlight. For example, this can be the rank for top-k facts, the percentage of a total sum for a mega-contributor, the percentage of dominated peers of a character for a peer-dominator, etc.

\silence{
A possible textual description of an Elementary Highlight is as follows:
\begin{quote}
	The character set $<CharacterSet>$ with the value $<Measure>$ = $<MeasureValue>$ serves as $<HighlightType>$ with a $<ScoreType>$ = $<ScoreValue>$.  
\end{quote}

As a side note, assuming a multidimensional space of $n$ dimensions, we can discriminate elementary highlights as (a) single-character elementary highlights, where the Character Set involves a single character, (b) point highlights, where there are exactly $n$ Characters from distinct dimensions, and, (c) the general case of multi-character highlights.
}  

\section{Discussion}\label{sec:discuss}
\paragraph{Who are the beneficiaries of the model?}
A first contribution of the proposed conceptual model is that it clarifies both the concepts and the terminology for data storytelling, for every stakeholder involved. Concerning \textit{data analysts}, the conceptual model allows the structuring of important parts of the problem in a way that is exploitable later: as soon as Highlight Types, Characters and important Measure Values become part of a structured solution, the data analyst can think on the problem in terms of them (e.g., ``what are the main Highlight Types hiding in my data? who are the main Characters in these data?"). Concerning \textit{tool builders}, it is absolutely feasible to direct the automation of algorithm execution and result structuring along the concepts of our model; once this automated extraction and representation of highlights is achieved, their exploitation for storytelling purposes is straightforward (see Fig.~\ref{fig:exTable01} and \ref{fig:exTable02}).

\paragraph{How realistic is the proposed model?}  Fig.~\ref{fig:exTable01} and \ref{fig:exTable02} report on a large number of typical data analysis algorithms. The fact that there is a straightforward translation to a structured result of all these heterogeneous algorithms that is immediately exploitable for data storytelling purposes testifies in favor of the  model proposed in this paper. For all practical purposes, the Holistic and Elementary Highlights proposed in this paper correspond to the results reported in the literature for tools automatically producing highlights. The data facts of  \cite{DBLP:journals/tvcg/WangSZCXMZ20,DBLP:journals/tvcg/ShiXSSC21} are Holistic Highlights; and the subject focus are Elementary Highlights. The data patterns of  \cite{DBLP:conf/sigmod/DingHXZZ19,DBLP:conf/sigmod/00040HZ21} are Holistic Highlights and the highlights are Elementary Highlights.\rem{drop a line in the RW that the discussion is here}

\paragraph{Implementation.} We have implemented our model in the Pythia data profiler \url{https://github.com/DAINTINESS-Group/Pythia}. Pythia automatically profiles the columns of submitted data sets for their descriptive stats, histograms, correlations, decision trees, outliers and dominance patterns. The tool extracts highlights along the lines of this paper too.

\silence{
\paragraph{Can we break free from the multidimensional spaces?}
Can we generalize from the multidimensional space setup to general data spaces?
The inherent advantage of such spaces is the integration of all the information in a single space of dimensions and measures (without the need to reconstruct data via joins of normalized tables), as well as the interweaving of different levels of abstraction in a single dimension. To be able to apply the model to arbitrary relational data spaces, 
we need the ability to automatically infer meaningful and useful join operations among different data sources in real time. Although the problem is clearly outside the scope of this paper, views, foreign keys and automatic extraction of functional dependencies in their absence can guide the process. However, the deep essence of phenomena and highlights is not affected by the nature of the data space.
}

\section{Conclusions}\label{sec:fw}
In this paper, we have presented a model that facilitates (a) the modeling, (b) the clarification of terminology, (c) the automation of the production of \textit{highlights}, i.e., \underline{structured} characterizations of subsets of the data space that are worth reporting due to their support of archetype statistical properties of interest, demonstrated as \textit{phenomena}. We have introduced the most important entities of the domain of highlight extraction and discussed their inter-relationships. \textit{Holistic highlights} are global properties that pertain to an entire set of facts, whereas \textit{elementary highlights} are their constituents that identify facts that play a particular role for the holistic highlight to take place. We have also demonstrated that (a) frequently encountered archetype properties are nicely covered by our modeling and (b) the highlight structure facilitates narratives straightforwardly.

In terms of future work, we are currently integrating the model into two tools concerning (a) automated query answering and (b) automated dataset profiling. The evaluation of highlight interestingness in a fully automated way, such that we can rank and prune highlights is an issue of open research questions. Structuring data stories in an efficient way, by taking advantage of the complementarity, overlap, discrepancy, or other properties of a set of highlights is another open research issue. 